\documentclass[useAMS,usenatbib,dvips]{mn2e}

\usepackage[dvips]{graphicx}            
\usepackage{pslatex}	    
\usepackage{graphicx}
\usepackage{color}
\usepackage[colorlinks=true, linkcolor=blue]{hyperref}

\title[Multifluid simulations of the Magnetorotational Instability in protostellar disks.]{Multifluid simulations of the Magnetorotational Instability in protostellar disks.}
\author[W. O'Keeffe and T.P. Downes]{W. O'Keeffe$^{1,3}$\thanks{E-mail:
wayne.okeeffe2@mail.dcu.ie (WOK)} and T.P. Downes$^{1,2,3}$\thanks{E-mail:
turlough.downes@dcu.ie (TPD)}\\
$^{1}$School of Mathematical Sciences, Dublin City University, Glasnevin, Dublin 9, Ireland\\
$^{2}$School of Cosmic Physics, Dublin Institute for Advanced Studies, 31 Fitzwilliam Place, Dublin 2, Ireland\\
$^{3}$National Centre for Plasma Science and Technology, Dublin City University, Glasnevin, Dublin 9, Ireland}
\begin{document}

\date{Accepted 2014 March 25. Received 2014 March 20; in original form 2013 September 18}

\pagerange{\pageref{firstpage}--\pageref{lastpage}} \pubyear{----}

\maketitle

\label{firstpage}

\begin{abstract}
Turbulent motion driven by the magnetorotational instability (MRI) is
believed to provide an anomalous viscosity strong enough to account for
observed accretion rates in protostellar accretion disks. In the first
of two papers, we perform large-scale, three fluid simulations of a
weakly ionised accretion disk and examine the linear and non-linear
development of the MRI in the net-flux and zero net-flux cases. This
numerical study is carried out using the multifluid MHD code HYDRA. We
examine the role of non-ideal effects, including ambipolar
diffusion, the Hall effect, and parallel resistivity, on the non-linear evolution of the MRI in weakly ionised protostellar disks in the region where the Hall effect is believed to dominate. 

We find that angular momentum transport, parametrised by the
$\alpha$-parameter, is enhanced by inclusion of non-ideal effects in the
parameter space of the disk model. The case where $\Omega \cdot
\mathbfit{B}$ is negative is explored and the Hall effect is shown to
have a stabilizing influence on the disk in this case.
\end{abstract}


\section{Introduction}

In protostellar accretion disks mass accretes onto the central object
with accretion rates in the range of 10$^{-7}$ - 10$^{-9}$
M$_{\odot}$\,yr$^{-1}$ \citep{Gullbring:1998}. For mass to accrete inwards radially, it is necessary for angular momentum to be transported outwards radially. A key question in accretion disk physics is what process or combination of processes is responsible for the efficient transport of angular momentum to obtain accretion rates in the observed range? The magnetorotational instability \citep{Balbus:1991} is the most promising transport mechanism proposed thus far.

The magnetorotational instability (MRI) has been studied extensively in the ideal MHD regime using local shearing-box
\citep{Hawley:1995, Stone:1996, Sano:2002a, Sano:2002b, Bai:2013a} and global \citep{Armitage:1998, Arlt:2001, Lyra:2008} numerical
simulations. Results obtained from these simulations have shown that the MRI process can lead to efficient transport of angular
momentum and may produce a rate of accretion similar to those obtained from observations. However, protostellar accretion disks are weakly ionised and so the ideal MHD regime may not be sufficient in describing the physics involved. Ionisation sources such as X-ray radiation from the protostar only ionize the surface of the disk, and accretion disks are typically too cold to provide effective thermal ionisation in the body of the disk. As a result, large regions adjacent to the mid-plane of the disk are not thought to be susceptible to the MRI \citep{Gammie:1996}, whereas the MRI is active closer to the surface of the disk.

In weakly ionised accretion disks, the plasma is largely made up from neutral fluid with a small fraction of the bulk fluid consisting
of a number of ionised species of differing properties. Three non-ideal processes are introduced by the interaction of the various
species, namely Ohmic dissipation, ambipolar diffusion and the Hall effect. In protostellar accretion disks, Ohmic dissipation is believed to dominate in high density regions with a low ionisation fraction such as the midplane and inner disk, ambipolar diffusion dominates in the opposite regime and the Hall effect is important in between \citep{Bai:2011b}. 

A number of studies have been performed where Ohmic dissipation is the dominant non-ideal process. When Ohmic dissipation is important
the MRI can be heavily suppressed. The importance of Ohmic dissipation can be measured by the magnetic Reynolds number, which
determines the relative importance advection and diffusion and is given by $R_M = c_s^2/ \eta \Omega$ where $\eta$ is the magnetic
diffusivity. A study by \citet{Fleming:2000} determined that the MRI cannot be self-sustained when $R_M \ge 10^4$.

Ambipolar diffusion arises from the imperfect coupling of the neutral and ionised fluids in a weakly ionised plasma such as that
typically found in accretion disks. The importance of ambipolar diffusion is determined by the value of the parameter $Am=\gamma
\rho_i/\Omega$ where $\rho_i$ is the density of the ionised fluid, $\gamma$ is the ion-neutral collisional coefficient, and $\Omega$
is the orbital frequency. It was determined in the linear analysis of \citet{Blaes:1994} that once the ion-neutral collision frequency
dropped below the orbital frequency, the MRI is suppressed. This result was confirmed by \citet{Hawley:1998} using numerical
simulations. In this case when $Am < 0.01$ the neutrals behave independently from the ions. In the opposite case, when $Am > 100$
ideal MHD behaviour is obtained. In the paper of \citet{Bai:2011a}, the above results were again confirmed using local simulations,
but it was seen that in the case of a net field it was possible for MHD turbulence to be maintained when $Am$ is small. The
effect of ambipolar diffusion is dependent on the field geometry. In local simulations such as the above, field geometry is
arbitrarily applied to the computational domain. The field geometry has important implications for the non-linear evolution of
the MRI and so to gain a deep understanding what effect ambipolar diffusion has on the saturation state of the MRI it is
necessary to use global simulations \citep{Bai:2011b}.
 
The Hall effect is very interesting as it can have a stabilising or destabilising effect depending upon the orientation of the magnetic
field because of the handedness introduced by the effect. We refer the reader to \citet{Wardle:2012} for details of the destabilisation mechanism. The linear analysis of \citet{Salmeron:2003} suggests that the Hall effect in this regime may have important consequences for the growth and saturation of the MRI. In the study of \citet{Sano:2002a, Sano:2002b}, the Hall-Ohm regime was investigated where Ohmic dissipation is large enough to have a damping effect on the MRI. Although the Hall effect was believed to allow the MRI to grow in situations where Ohmic and ambipolar diffusion would normally suppress the MRI entirely \citep{Wardle:1999}, the simulations presented in this study did not investigate this specific point. However the parameter space investigated was narrow and did not include the Hall dominated case which would be common in protostellar disks where a large extent of the disk may be in this regime. In this regime it is believed that estimates of the depth of magnetically active layers in accretion disks may be inaccurate because of disk destabilisation due to the Hall effect and protostellar disks may be turbulent much closer to the mid-plane of the disk \citep{Wardle:2012}.

The scenario where the Hall effect dominates over the other non-ideal processes has only recently begun to be explored. In the paper of \citet{Kunz:2013}, the saturation of the MRI in a weakly ionised protostellar disk, dominated by the Hall effect, is studied using the shearing box 
approximation. The authors describe "large scale" axisymmetric effects (insofar as local simulations can make predictions about large scale effects) forming in the magnetic and velocity fields. The structures described by the authors are very long-lived and have the net effect of reducing the amount of MRI driven turbulent transport present. These results suggest that the Hall effect, may not be as effective in allowing the MRI to develop in regions which should be inactive to MRI driven turbulence as suggested by \citet{Wardle:2012}. Their results also appear to be in contradiction to those of \citet{Bai:2013a} and \citet{Salmeron:2003,Salmeron:2005}, although the authors acknowledge that the inclusion of a stratification may be important. The model of \citet{Kunz:2013} occupies a similar parameter space to the global multifluid model presented here.

In this paper, we present the first fully multifluid MHD, global numerical simulations of accretion disks in a region of the disk where the Hall effect is believed to dominate over other non-ideal effects. In the presence of purely diffusive non-ideal effects, angular momentum transport would be expected to be damped compared to the ideal MHD case. This work investigates the destabilising nature of the Hall effect and illustrates the importance of a net magnetic field and its orientation relative to the net angular momentum vector in determining the non-linear evolution of the disk in the presence of the Hall effect, confirming the results of \citep{Sano:2002a, Sano:2002b}.

The structure of this paper is as follows. In section 2 we present the numerical set-up and explain the model and boundary conditions used. In section 3, the analysis techniques used to produce the results are explained. Section 4 contains a validation of the model. In section 5, a resolution study is performed and the results of a direct comparison between the ideal MHD and fully multifluid models are presented. Finally, we conclude and discuss the implications of the results in section 6.         

\section[]{Numerical Approach}

The numerical simulations described in this paper are performed using the multifluid magnetohydrodynamic code HYDRA which models
weakly ionised plasmas \citep{OSullivan:2006, OSullivan:2007}. The multifluid simulations include three fluids, consisting of a neutral species, an electron and an ion species.

\subsection{Multifluid equations}
The following set of equations model the evolution of a neutral fluid and $N-1$ charged fluids in the presence of a magnetic field.
These equations govern a weakly ionised plasma. In the case of a weakly ionised plasma, the pressure and inertial terms of the
evolution equations for the ionised species (subscripted by $i$) may be ignored. This is because the mass density of the plasma is dominated 
by the neutral component \citep[e.g.][]{Falle:2003, Ciolek:2002}. An isothermal closure relation is used, therefore allowing the energy equations for the neutrals and ionised species to be ignored. 

\begin{equation}
\frac{\partial \rho_{i}}{\partial t} + \nabla \cdot (\rho_{i}  \mathbfit{v}_{i}) = 0
\end{equation}

\begin{equation}
\frac{\partial \rho_{1} \mathbfit{v}_{1}}{\partial t} + \nabla \cdot (\rho_{1} \mathbfit{v}_{1}\mathbfit{v}_{1} + p_{1} \mathbfss{I} ) =
\mathbfit{J} \times \mathbfit{B}
\end{equation}

\begin{equation}
\frac{\partial \mathbfit{B}}{\partial t} + \nabla \cdot (\mathbfit{v}_{1} \mathbfit{B} - \mathbfit{B} \mathbfit{v}_{1} ) = - \nabla
\times \mathbfit{E}^{'}
\end{equation}

\begin{equation}
\alpha_{i} \rho_{i}(\mathbfit{E} + \mathbfit{v}_{i} \times \mathbfit{B}) + \rho_{i} \rho_{1} K_{i1} (\mathbfit{v}_{1} - \mathbfit{v}_{i}) = 0 \: ; \: (2 \leq i \leq N) 
\end{equation}

\begin{equation}
\nabla \cdot \mathbfit{B} = 0
\end{equation}

\begin{equation}
\nabla \times \mathbfit{B} = \mathbfit{J}
\end{equation}

\begin{equation}
\displaystyle\sum_{i=2}^{N} \alpha_{i} \rho_{i} = 0
\end{equation}

\begin{equation}
\displaystyle\sum_{i=2}^{N} \alpha_{i} \rho_{i} \mathbfit{v}_{i} = \mathbfit{J}
\end{equation}

The subscripts in the above equations denote the species in question. The neutral component is represented by a subscript of 1, and
the charged species are denoted by subscript 2 to $N$. The variables $\rho_{i}$, $\mathbfit{v}_{i}$, $p_{i}$ represent the mass density, pressure and velocity of species $i$ respectively. The collisional coefficient $K_{i1}$ represents the collisional interaction
between species $i$ and the neutral fluid. The charge to mass ratio of the $i^{\rm th}$ species is represented by $\alpha_{i}$.
The identity matrix, current density, and magnetic flux density are represented by \textbfss{I}, \textbfit{J} and \textbfit{B}
respectively and finally the full electric field is given by $\mathbfit{E}=-\mathbfit{v}_{1} \times \mathbfit{B} + \mathbfit{E}^{'}$. One more equation is needed to close the set. An equation of state may be used. In this case the isothermal relation $c_{s}^{2}=p_{1}/\rho_{1}$ is added to the set.

This set of equations leads to an expression for the electric field in the frame of the fluid, $\mathbfit{E}^{'}$, given by the generalised Ohm's law

\begin{equation}
\mathbfit{E}^{'} = \mathbfit{E}_O + \mathbfit{E}_H + \mathbfit{E}_A 
\end{equation}

\noindent where the electric field components are given by

\begin{equation}
\mathbfit{E}_O = (\mathbfit{J} \cdot \mathbfit{a}_O)\mathbfit{a}_O 
\end{equation}

\begin{equation}
\mathbfit{E}_H = \mathbfit{J} \times \mathbfit{a}_H 
\end{equation}

\begin{equation}
\mathbfit{E}_A = -(\mathbfit{J} \times \mathbfit{a}_H) \times \mathbfit{a}_H 
\end{equation}

\noindent using the definitions $\mathbfit{a}_O \equiv f_O \mathbfit{B}$, $\mathbfit{a}_H \equiv f_H \mathbfit{B}$ and $\mathbfit{a}_A \equiv f_A \mathbfit{B}$ where $f_O \equiv \sqrt{r_O}/B$, $f_H \equiv r_H/B$ and $f_A \equiv \sqrt{r_A}/B$.

The resistivities are given by

\begin{equation}
r_{o} \equiv \frac{1}{\sigma_o} 
\end{equation}

\begin{equation}
r_{H} \equiv \frac{\sigma_{H}}{\sigma_{H}^{2} + \sigma_{A}^{2}} 
\end{equation}

\begin{equation}
r_{A} \equiv \frac{\sigma_{A}}{\sigma_{H}^{2} + \sigma_{A}^{2}} 
\end{equation}

\noindent where the conductivities are defined for each charged species $i$ by

\begin{equation}
\sigma_{o} = \frac{1}{B} \displaystyle\sum_{i=2}^{n} \alpha_{i} \rho_{i} \beta_{i}
\end{equation}

\begin{equation}
\sigma_{H} = \frac{1}{B} \displaystyle\sum_{i=2}^{n} \frac{\alpha_{i} \rho_{i}} {1 + \beta_{i}^{2}}
\end{equation}

\begin{equation}
\sigma_{A} = \frac{1}{B} \displaystyle\sum_{i=2}^{n} \frac{\alpha_{i} \rho_{i} \beta_{i}} {1 + \beta_{i}^{2}}
\end{equation}

where ($\beta_{i}$) is the Hall parameter for the charged species and is a measure of how well tied to the magnetic field the charged species ($i$). This is given by

\begin{equation}
\beta_{i} = \frac{\alpha_{i} B}{K_{i1} \rho_{1}}
\end{equation}

To solve these equations numerically, the time integration of these equations is multiplicatively operator split into 3 separate operations which may be summarised as follows:

\begin{enumerate}

\item {The neutral fluid is advanced. Equations (1), (2) and (3), with index $i=1$ for equation (1) are solved using a standard finite volume integration method to 2nd order temporal and spatial accuracy. The diffusive term on the RHS of equation (3) is not evaluated until a later step. The restriction described by equation (5) is also maintained using the method of \citet{Dedner:2002}.}

\item{The diffusive term in equation (3) is now evaluated. By using
	standard discretisation, a restriction is imposed by the potentially vanishing timestep associated with very high values of Hall resistivity \citep{Falle:2003}. To overcome this problem, the induction equation is integrated by multiplicatively operator splitting the Hall and ambipolar terms. Special techniques, known as super-timestepping and the Hall diffusion scheme are then used to advance the Hall and ambipolar terms efficiently while simultaneously relaxing the timestep restriction and maintaining 2nd order accuracy.}

\item{Finally, the densities and velocities of the various charged species are evaluated. The densities are evaluated using the mass conservation equation (1) for $2 \leq i \leq N$, where $N$ is the number of fluids.}
\end{enumerate}

For further details on the numerical code HYDRA, including discussions of stability and accuracy of the results, we refer the reader to \citet{OSullivan:2006,OSullivan:2007}.

\subsection{The model}
The computational model used in this study is similar to that found in \citet{Lyra:2008}. A Cartesian grid is used in this model. This approach has been used for accretion disk studies before with success \citep{Peplinski:2008,Zhang:2008}. It has been shown rigorously by \citet{Valborro:2006} that codes based on Cartesian grids produce results that are not significantly different to those produced with other grid types.

To save computational expenditure, we only simulate one quadrant of the accretion disk \citep{OKeeffe:2013}. This approach has been used in the accretion tori study of \citet{Stone:2000} where it was found that using a $\pi/2$ domain does not significantly change the quantitative results at saturation. It is possible to further reduce the azimuthal domain \citep{Dzyurkevich:2010}, however as the domain is reduced the results will begin to diverge from those obtained using a complete azimuthal domain \citep{Flock:2012}. This occurs because the larger azimuthal modes of the MRI no longer fit within the domain. The use of a $\pi/2$ domain is convenient when using a Cartesian coordinate system where a smaller azimuthal domain would be non-trivial to set-up.

\subsection{Boundary conditions}
Implementing a global accretion disk model on a Cartesian grid presents problems with how to deal with the boundaries. The boundary conditions can be categorised as box-face and interior boundaries. The box-face boundaries are applied to the sides of the computational box in which the disk is situated. The interior boundaries are placed inside this box to take into account the cylindrical nature of the disk. Outside of the cylindrical disk, the fluid variables are not allowed to evolve and are essentially frozen. The fluid is allowed to flow in and out of these interior boundaries and so some mass and magnetic energy is expected to flow through these boundaries and the disk is not a closed system. These two types of boundaries will now be described in turn.

\begin{figure}
\includegraphics[width=84mm]{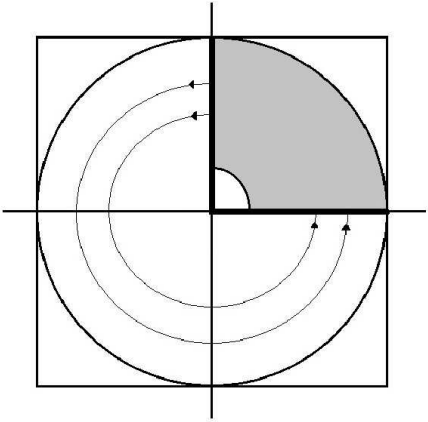}
\caption{Illustration of computational domain and boundary conditions.
The upper right hand quadrant is the extent of the Cartesian grid. The thickened
edges of this quadrant represent the azimuthal boundaries. Material which leaves 
through the low-$x1$ boundary will re-enter through the low-$x2$ boundary and 
vice-versa. The arrows illustrate this. The shaded region in the upper right hand 
quadrant is the active computational domain and the white regions are the frozen zones.
The line between the active and frozen domain is the wavekilling interior boundary.
}
\end{figure}

\subsubsection{Box-face boundaries}
The box-face boundaries are applied to the sides of the computational
box. The upper and lower $z$ boundaries are simply periodic as we assume there is no vertical gravity gradient. This ensures that no mass or energy is lost through the vertical boundaries. The upper $x$ and $y$ boundaries are unimportant as they are directly adjacent to the frozen zone. No fluid crosses these boundaries and they can simply be ignored. The lower $x$ and $y$ boundaries are periodic with each other. Fluid that flows out of one boundary is transported across the other. This allows and is consistent with the use of the quarter disk approximation mentioned at the beginning of this section.

\subsubsection{Interior boundaries}
The accretion disk itself is a quarter cylinder which is placed in a flattened square prism. This sets a limit to the radial extent of the disk i.e. $r\leq L_{x}$. This condition makes it necessary to introduce an interior radial boundary. As mentioned previously, a frozen region is introduced for $r \leq r_{\rm{int}}$ and $r \geq r_{\rm{ext}}$ where the dynamical equations are not evolved for the fluids or the magnetic field. In the models presented here $r_{\rm{int}} = 0.5$, $r_{\rm{ext}} = 2.58$ and $L_x = L_y = 2.6$ and $L_z=0.075 \times L_x$ . The inner bufferzone is positioned just outside the inner frozen zone at $0.5 < r < 0.6$, similarly the outer bufferzone is positioned at $2.48 < r < 2.58$. Fig.2 shows the position of these zones relative to the box boundaries. The purpose of the radial buffer zone is to smoothly transition the fluid parameters so that any numerical instabilities associated with the abrupt jump from free to frozen regions may be avoided \citep{Lyra:2008}. There is a dual purpose to this type of inner boundary. It allows an escape route for undesirable waves at the beginning of the simulation. In the models presented here, transient waves are seen to originate from the frozen region at the beginning of the simulations. If the wavekilling boundaries were not present these waves could reflect back and forth throughout the computational domain. It is important to note that the bufferzones are not fixed in nature and allow accretion through them.  

\begin{figure}
\includegraphics[width=84mm]{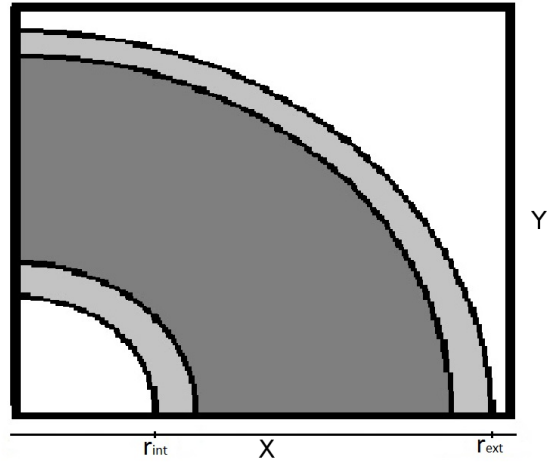}
\caption{Schematic showing the x,y plane of the computational domain.
The dark shaded area is the active computational domain. The white areas
bounded by the black curves and the sides of the box are the frozen regions.
The light shaded regions are the wavekilling interior boundaries. Refer to
section 2.3.2 and Fig.1 for a detailed description.
}
\end{figure}

The inner buffer zone must kill waves faster as the fluid evolves
dynamically on shorter time-scales here and hence the risk of
instabilities arising at this boundary is higher. It is worth noting that \citet{Lyra:2008} performed some simulations without this inner boundary and models with and without inner radial boundaries behave similarly with little quantitative difference. A smaller timestep results as the fluid will be rotating at a much higher rate closer to the origin than at the inner radial edge of the bufferzone.

Following \citet{Lyra:2008}, the bufferzone drives the fluid parameter $X$ gradually and smoothly to its initial condition such that

\begin{equation}
\frac{dX}{dt}= -\frac{X-X_0}{\tau }S(r)
\end{equation}  

\noindent where $S(r)$ is the driving function which has a range of [0,1.0], and $\tau$ is the orbital period at the boundary in question. Linear and non-linear functions were evaluated for the driving function. It was decided to use a linear driving function as no quantitative difference was found between the linear and non-linear functions. The driving function is defined by the following piecewise linear function:

\[S(r) = \left\{
  \begin{array}{ll}
    1.0 & \textrm{if } r \leq 0.5\\
    10(0.6-r) & \textrm{if } 0.5 < r < 0.6\\
    0 & \textrm{if } 0.6 \leq r \leq 2.48\\
    10(r-2.48) & \textrm{if } 2.48 < r < 2.58\\
    1.0 & \textrm{if } r \ge 2.58
  \end{array}
\right.
\] 

\subsection{Gravity}
A Newtonian gravitational potential is applied to the fluid within the computational domain

\begin{equation}
\Phi = - \frac{GM}{\sqrt{x^2 + y^2}+c}
\end{equation}

\noindent where $G$ is the universal constant of gravitation, $M$ is the mass of the central object, and c is a softening parameter. Notice that this gives a cylindrical potential, where the fluid is attracted to a pole whose centre is located at $x=y=0$ rather than a single point. Whereas normally a radial and vertical stratification would exist, in this case vertical stratification is omitted. The vertical extent of the disk model is small compared to the radial domain size but large enough that vertical structure in the densities and magnetic field may develop without influence from the vertical boundaries.    

\subsection{Initial conditions}
For the ideal MHD model the following initial conditions are used. The neutral density is uniform initially and is set to 1.17$\times$10$^{-11}$g/cm$^3$. The average mass of the neutral particles is taken to be $m_{\rm{n}} = 2.33 m_{\rm{p}}$ , where m$_{\rm{p}}$ is the mass of a proton. This corresponds to a fluid of 90\% molecular hydrogen and 10\% atomic helium by number, which is representative of molecular clouds. There is no pressure gradient initially. All models presented are isothermal and the neutral pressure is found using the isothermal relation $a=\sqrt{P/\rho}$.

The choice of an appropriate model for the study of the MRI is not trivial. From linear analysis \citep{Balbus:1991} of the MRI it is found that only a weak magnetic field, and differential rotation is required for the MRI to be present. In numerical simulations, it is necessary that the processes involved in an instability be adequately resolved by the computational grid. In the case of the MRI, it is found from linear analysis that a critical vertical wavelength must be resolved \citep{Balbus:1991, Balbus:1998}:

\begin{equation}
\lambda_{c}=\frac{2\pi}{\sqrt{3}} \frac{v_{A}}{\Omega}
\end{equation} 

\noindent where $v_{A}$ is the Alfv\'{e}n speed which is given by

\begin{equation}
v_{Az}=\frac{B_z}{\sqrt{\mu_{0} \rho}}
\end{equation}

The critical wavelength of the MRI is approximately the distance an Alfv\'{e}n wave would travel vertically in a single orbital period. The model must be constructed while considering the requirement that this critical wavelength be resolved either initially or become resolved though the use of perturbation at an appropriate time. An initial magnetic flux of 50mG is chosen. Magnetic fields are known to reach a strength of an order of a 1G in the region of interest in this work \citep{Konigl:1993}. 

The critical wavelength of the MRI using this initial value of magnetic flux is resolved by 12 cells at the $r=2.0$ when using the finest resolution. At $r=1.0$ the critical wavelength is marginally resolved by approximately 4 zones but becomes resolved soon after the fluid perturbation is applied. Using the criterion of \citet{Hawley:2013}, the critical wavelength is sufficiently resolved at the outer radii, however after the fluid perturbation is applied, the inner radii begin to succumb to MRI driven turbulence and the MRI slowly becomes better resolved. At saturation, the $\lambda_z$ is resolved by 12-15 cells and $\lambda_\phi$ is resolved by 25-30 cells at $r=1.0$. 

The densities for the charged species are much smaller than the neutral species as the plasma is weakly ionised. The ion fluid represents an average of ions produced from a number of metal atoms, including Na, Mg, Al, Ca, Fe and Ni. These metals have sufficiently similar ionisation and recombination rates, and so can be modelled collectively as ions of a single positive charge \citep{Umebayashi:1990}. The molecular ions, of which HCO$^+$ is the most numerous, are significantly less abundant than the metal ions, allowing us to neglect them. An average mass of $m_{\rm{ion}} = 24 m_{\rm{p}}$ is assigned to the particles of the ion fluid, approximately equal to that of a magnesium ion.

An ionisation fraction of approximately 4$\times 10^{-11}$ is expected
in the radial range studied in this work. The number density of the
neutral fluid is expected to be approximately 7$\times 10^{12}$cm$^{-3}$
\citep{Salmeron:2003}. The ionisation fraction ($\zeta$) is given by
$\zeta=n_{\rm{e}}/n_{\rm{H}}$ leading to a number density for the electron fluid of approximately 300 cm$^{-3}$. 

The charge to mass ratios are calculated as follows,

\begin{equation}
\alpha_{\mathrm{ion}} = \frac{+e}{m_{\mathrm{ion}}}=\frac{1.6 \times 10^{-19}\mathrm{C}}{24m_{\mathrm{p}}}=4.0\times 10^{3} \mathrm{g}^{-1}=1.2\times 10^{13} \mathrm{stat C \, g}^{-1}
\end{equation}  

\begin{equation}
\alpha_{\mathrm{e}} = \frac{-{e}}{m_{\mathrm{e}}}=\frac{1.6 \times 10^{-19}\mathrm{C}}{m_{\mathrm{e}}}=-5.27 \times 10^{17} \mathrm{stat C \, g}^{-1}
\end{equation}

The number density of the ion fluid is then found using the charge neutrality condition
\begin{equation}
\alpha_{\mathrm{ion}} \rho_{\mathrm{ion}} + \alpha_{\mathrm{e}} \rho_{\mathrm{e}} = 0
\end{equation}
\noindent The ion density, $\rho_{\rm{ion}}$, is then found to be 1.2$\times10^{-20}$ g cm$^{-3}$.

The importance of the multifluid effects are set through the collision
coefficients. Collisions between the charged fluids are not considered
as their respective densities are low and so such collisions are rare in
comparison to the collisions between charged particles and neutrals. The rate coefficient for the ion fluid may be found in \citet{Wardle:1999}. The collisional coefficients are found thus

\begin{equation}
K_{\mathrm{ion,n}} = \frac{<\sigma \nu>_{\mathrm{ion}}}{m_{\mathrm{ion}}+m_n} = \frac{1.6 \times 10^{-19} \mathrm{cm}^3 \mathrm{s}^{-1}} {24m_{\mathrm{p}}+2.33m_{\mathrm{p}}} = 3.64 \times 10^{13} \mathrm{cm}^3 \mathrm{g}^{-1} \mathrm{s}^{-1}
\end{equation}

\begin{equation}
K_{\mathrm{e,n}} = \frac{<\sigma \nu>_{\mathrm{e}}}{m_{\mathrm{e}}+m_{\mathrm{n}}} = \frac{1.15\times^{-15}\left( \frac{128 K_{\mathrm{B}} T_{\mathrm{e}}}{9\pi m_{\mathrm{e}}} \right)}{m_{\mathrm{e}} + 2.33m_{\mathrm{p}}}=2.88 \times 10^{15} \mathrm{cm}^3 \mathrm{g}^{-1} \mathrm{s}^{-1}
\end{equation}

\section{Analysis}

A number of parameters are calculated to compare the solutions of the simulations. These will now be detailed. 

The stresses associated with the differential rotation, that is typical in accretion disks, are commonly studied parameters. These are the Reynolds (fluid) and Maxwell (magnetic) stresses. The Maxwell stress is the most important in the study of the MRI as the magnetic stress mediates the MRI process. In accretion disks, both wavelike and turbulent disturbances can create tight radial-azimuthal correlations in the fluctuations in the velocity and magnetic fields \citep{Balbus:2003}, the $r-\phi$ component of the stress is associated with the viscous torque that provides the angular momentum transport, therefore the $r-\phi$ component of the stresses is an important parameter to observe in both local and global simulations involving the MRI and give a good indication as to the magnitude of angular momentum transport present. The $r-\phi$ component of the stresses may be represented together as the sum of the Reynolds and Maxwell stresses

\begin{equation}
T_{r \phi} = \overline{\rho \delta V_r \delta V_{\phi}} - \overline{\delta B_r \delta B_{\phi}}
\end{equation}

\noindent where T$_{r \phi}$ is the $r-\phi$ component of the total stress tensor, and $\delta B$ denotes the fluctuating component of the magnetic field and is given by

\begin{equation}
\delta \mathbfit{B}_x = \mathbfit{B}_x - \overline{\mathbfit{B}_x}
\end{equation}

\noindent where $x$ represents either the radial or azimuthal component of the magnetic field.

In the case of the fluctuating part of the velocities ($\delta v$), the
Keplerian velocity field at the radius at which the sampled fluid
element is found to be ($\overline{\mathbfit{v}_k}$) is first subtracted and then the fluctuating component is found to give

\begin{equation}
\delta \mathbfit{v}_x=\mathbfit{v}_x - \overline{\mathbfit{v}_k}
\end{equation}

Another important quantity that is used in MRI studies is the anomalous viscosity which is parametrised by the $\alpha$-parameter \citep{Shakura:1973} where

\begin{equation}
\alpha = \alpha_{R} + \alpha_{M}
\end{equation}

\noindent where $\alpha_{R}$ and $\alpha_{M}$ are the kinetic and magnetic components of the anomalous viscosity and are given by

\begin{equation}
\alpha_{R} = \frac{\overline{\rho \delta V_r \delta V_{\phi}}}{\rho c_{s}^2}
\end{equation}

\begin{equation}
\alpha_{M} = -\frac{\overline{ \delta B_r \delta B_{\phi}}}{\rho c_{s}^2}
\end{equation}

\noindent where $c_s$ is the isothermal sound speed.

When calculating the stresses, a small region adjacent to the inner and
outer boundaries is ignored. The inner boundary conditions are seen to
directly effect the fluid flow locally. As a test of our results this
boundary was removed in one simulation and the evolution of the
properties of the simulation in which we are interested was shown to be consistent with a simulation with the inner boundary included. The same could not be done with the outer boundary due to the fluid simply flowing out of the grid if the frozen zone is not present.

Other parameters that are useful to compare are the temporal evolution
of the magnetic and kinetic energies. The total kinetic energy itself is
not particularly useful apart from showing that the total energy of the
system does not change much over the time simulated. Some change is to
be expected since the choice of boundary conditions allow inflow and
outflow of mass and kinetic and magnetic energy. In practice however, the flow on to the grid closely matches the outflow of mass and energy out of the grid and so the increase/decrease of mass and energy over the lifetime of these simulations is very small. Instead, the perturbed kinetic energy is used for comparison purposes and is given by

\begin{equation}
E_k = \int \int \int \,\frac{1}{2} \rho \,[\delta \mathbfit{v}^2 - \mathbfit{v}_{k}^2] \, dx \, dy \, dz 
\end{equation}

\noindent where $\mathbfit{v}_k$ is the Keplerian velocity.

The total magnetic energy is important to follow as it is expected to grow as the stretching and twisting of the magnetic field due to fluid motions extracts kinetic energy from the flow. The total magnetic energy is given by

\begin{equation}
E_M = \int \int \int \frac{1}{2}\mathbfit{B}^2 \, dx \, dy \, dz
\end{equation}

\section{Validation}
To validate our approach, a comparison is made between the global and quarter disk models using the analysis techniques outlined above using the ideal MHD model. The set-up for the global model is identical to that described in the previous section, the only difference is that the azimuthal domain is 2$\pi$ radians. The res4-ideal model is compared to a model with a grid size of 720x720x27 with the disk center set at the center of the domain at a vertex. 

The evolution of the stresses and energies are computed for 70 orbital periods which is longer than the time period used in production runs (see table 1). Some divergence between the results occur towards the end of the simulations in the total magnetic energy. Qualitatively, this is seen in the region nearest the inner boundary in the global model and this difference is caused by a build-up of accreted magnetic energy at the inner boundary between the fixed and free regions which is not a severe as the quarter disk case but this does not occur until well past the typical end point of our production runs. Diagnostics are performed slightly away from the inner boundaries to make sure this effect does not cause spurious results in production. 

\begin{figure}
\includegraphics[width=84mm]{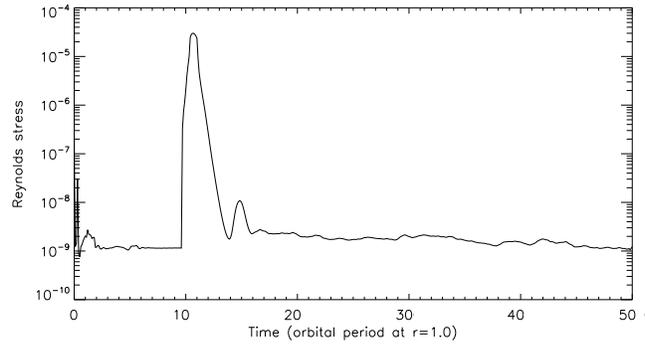}
\caption{Plot of the Reynolds stress for the hydro run. The fluctuations in the disk can be seen early on and soon settle down. The perturbation is injected after 10 orbits and the disk quickly settles down to it's initial state.
}
\end{figure}

A hydrodynamical simulation was also run to show that no turbulence exists due to hydrodynamical instabilities. This run is identical to the model described above but the magnetic field is not present. In this case the perturbation is seen to quickly die out and the disk returns to a stable state (see Fig.3). No evidence is seen of any sustained turbulence on the grid. The accretion rate is calculated for the hydrodynamical simulation and is compared to res4-ideal 
(see Fig.4).  As can be seen, virtually no accretion occurs, implying
that virtually no transport of angular momentum occurs, in the
hydrodynamic case.  This supports, for this work, the general
conclusions of \citet{Valborro:2006} regarding the appropriateness of
Cartesian grids for simulations of accretion disks.

\begin{figure}
\includegraphics[width=84mm]{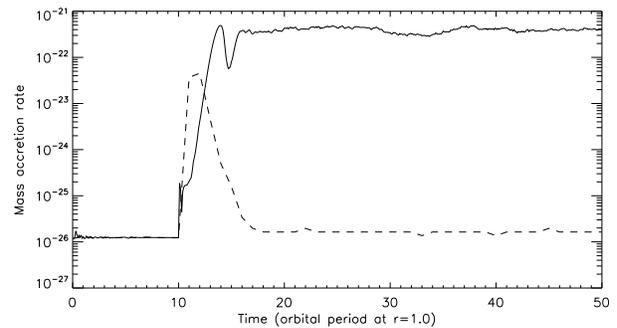}
\caption{Plot of the mass accretion rate per cm height of the cylindrical disk (M$_{\odot}$\,yr$^{-1}$cm$^{-1}$) against time for the hydro (dashed line) run and res4-ideal (solid line) run. Mass accretion is seen in the hydro run related to the injection of the fluid perturbation but this quickly dies away.
}
\end{figure}

\section{Results}
The results are presented in the following two sections. Firstly, a resolution study is carried out for the ideal MHD and multifluid models to ensure that the solution converges adequately. Once a final resolution has been decided upon, a direct comparison will be made of the ideal MHD and multifluid regimes.

\subsection{Resolution study}

A series of simulations are performed where cell spacing is decreased, starting with the coarsest resolution and ending with the finest. At the finest resolution, the critical vertical wavelength of the MRI is initially resolved by 12 zones. The parameters used for the comparison of the various runs are calculated from the point of saturation for the same period of evolution (see table 1). The point at which growth ceases is different for each resolution studied, what is important is that the mean and error calculations are taken over the same time periods. The error is given by the standard deviation over the same time period.

As expected in the ideal MHD case the solution converges asymptotically
so that the difference between the res4-ideal and res3-ideal runs is
less than 10\% for all parameters. Interestingly the multifluid case
seems to be converging but the finest resolution case diverges
significantly. 

\begin{table*}
\caption{Resolution study and production run data.}
\begin{tabular}{@{}lcccccc}
\hline
Run ID & $N_{x,y}$ & $N_{z}$ & $\alpha$ & $-M_{r \phi}$ & $E_M$     & $E_K$     \\
       &           &         & $10^{-2}$& $10^{-5}$     & $10^{-4}$ & $10^{-5}$ \\
\hline
Global-ideal & 720 & 27  & 2.72 $\pm$ 0.37 & 4.3024 $\pm$ 0.5239  & 3.4152 $\pm$ 0.6912  & 2.3906 $\pm$ 0.3187 \\
Res1-ideal & 120 & 9  & $\approx$0.0    & $\approx$0.0         & $\approx$0.0         & $\approx$0.0        \\
Res2-ideal & 240 & 18 & 1.18 $\pm$ 0.23 & 2.5673 $\pm$ 0.4365  & 2.489 $\pm$ 0.7976   & 0.9869 $\pm$ 0.0253 \\
Res3-ideal & 360 & 27 & 2.54 $\pm$ 0.33 & 3.9488 $\pm$ 0.6934  & 3.360 $\pm$ 0.7121   & 2.4471 $\pm$ 0.3722 \\
Res4-ideal & 480 & 36 & 2.80 $\pm$ 0.4  & 4.3085 $\pm$ 0.7583  & 3.084 $\pm$ 0.5719   & 2.7598 $\pm$ 0.2799 \\
Res1-mf & 120 & 9  & $\approx$0.0        & $\approx$0.0        & $\approx$0.0         & $\approx$0.0        \\
Res2-mf & 240 & 18 & 1.56 $\pm$ 0.43 & 3.1177 $\pm$ 1.2139 & 4.6600 $\pm$ 0.8784  & 0.9684 $\pm$ 0.1428 \\
Res3-mf & 360 & 27 & 2.65 $\pm$ 0.22 & 4.9330 $\pm$ 0.4634 & 3.7673 $\pm$ 0.6756  & 1.9608 $\pm$ 0.0088 \\
Res4-mf & 480 & 36 & 7.52 $\pm$ 0.42 & 12.987 $\pm$ 0.7558 & 0.002764 (Max.)      & 6.8454 $\pm$ 0.4838 \\
480-mf-minusbz & 480 & 36 & 0.92 $\pm$ 0.035 & 1.1243 $\pm$ 0.6218 & 0.922 $\pm$ 0.0497   & 1.8740 $\pm$ 0.3398 \\
480-mf-zeronet & 480 & 36 & 0.45 $\pm$ 0.041 & 0.7722 $\pm$ 0.0516 & 0.804 $\pm$ 0.0299   & 6.8454 $\pm$ 0.5194 \\
\hline
\end{tabular}
\end{table*}

\begin{figure}
\includegraphics[width=84mm]{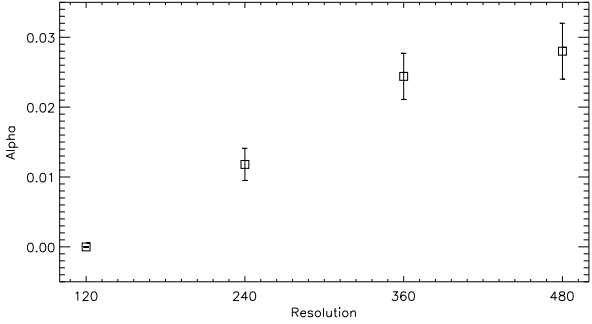}
\caption{Plot of the $\alpha$-parameter against resolution for the ideal MHD resolution study.
}
\end{figure}

\begin{figure}
\includegraphics[width=84mm]{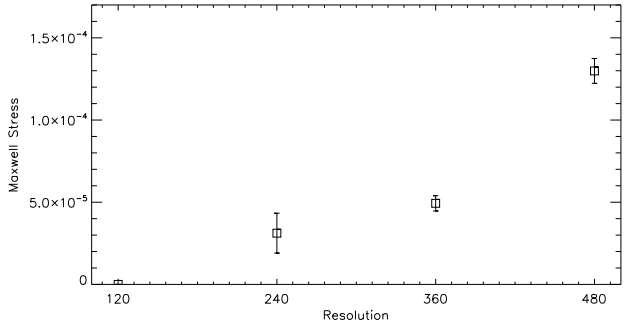}
\caption{Plot of the Maxwell stress against resolution for the multifluid resolution study.
}
\end{figure}

From the initial conditions it is possible to get an estimate of the length scales for the Hall effect and ambipolar diffusion and compare these to the characteristic lengthscale of the system.

The Hall term becomes important when the ion inertial length becomes comparable to the characteristic length scale given by

\begin{equation}
	L_H = \frac{c}{\omega_{pi}} \frac{v_A}{V_0}
\end{equation}

\noindent where $\omega_{pi}$ is the ion plasma frequency, $c$ is the speed of light, $v_A$ is the Alfv\`{e}n speed given by $v_A=B/\sqrt{4 \pi n_i m_i}$, and $V_0$ is the characteristic speed of the flow which in this case is the orbital speed \citep{Mininni:2003}.

Using the initial conditions at r=2.0, the Hall scale ($L_H$) is calculated to be 8.89$\times 10^{8}$cm. In comparison, the cell spacing in the simulations vary from approximately 0.1 to 0.03 AU or 1.62$\times 10^{12}$cm to 4.05$\times 10^{11}$cm. It is clear that the Hall scale is much less than the grid scale initially. However, once the MRI begins to become active the magnetic field strength will amplify, also turbulence in the ionised species causes the ion density to have a wide distribution about the mean ion density and so locally the Hall scale can and does reach the same order of magnitude of the grid scale.

In the case of ambipolar diffusion, typical length scales are approximately given by

\begin{equation}
L_{AD}=\frac{B^2}{4 \pi \rho_i \rho_n K_{ion,n} V_0}
\end{equation}

\noindent where $\rho_i$ and $\rho_n$ are the ion and neutral fluid densities respectively, $K_{ion,n}$ is the ion-neutral collisional frequency and $V_0$ is the characteristic velocity of the flow \citep{Oichi:2006}. Initially, as in the Hall case, the length scale associated with ambipolar diffusion is much smaller than the grid scale. At r=2.0, $L_{AD}$ is calculated to be 6.18$\times 10^8$cm using the initial conditions. By examining the properties of regions of the disk at the same radius once the MRI has saturated, the ambipolar length scale is found to be in the range 0.01 $\Delta x$ $\leq$ $L_{AD}$ $\leq$ $ \Delta x$. 
 
\subsection{Comparison of ideal MHD and multifluid regimes}
It is obvious from table 1 that the ideal MHD and multifluid simulations differ significantly. It might be expected that diffusive processes would lead to a suppression of the MRI but the destabilising of the disk through the Hall effect leads to larger $\alpha$-parameter. In this section, the stresses and energies for the res4-ideal and res4-mf runs are compared.
 
\subsubsection{Stresses and $\alpha$-parameter}

The stresses show a considerable difference between the ideal MHD and multifluid cases (see Fig.7 and Fig.8). In the ideal MHD regime (res4-ideal), after the initial growth, the Maxwell stress and $\alpha$-parameter show little change at saturation. In the multifluid (res4-mf) case, the Maxwell stress and $\alpha$-parameter appear to begin to saturate at similar levels as seen in the res4-ideal run but then a renewed period of growth which appears to be linear occurs, which eventually saturates at a level roughly 2.5 times that of the res4-ideal run (see Table 1). This suggests that considerable accretion is occurring in the res4-mf run.

\begin{figure}
\includegraphics[width=84mm]{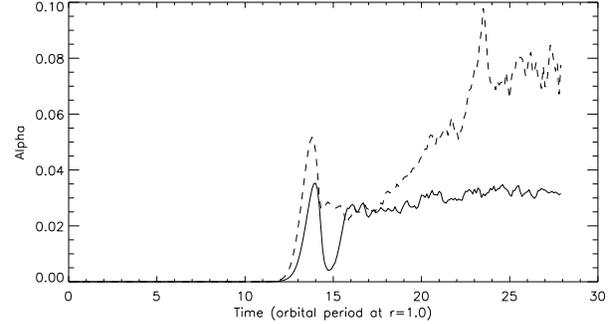}
\caption{Plot of the ideal MHD (solid line) and multifluid (dashed line) $\alpha$-parameter.
}
\end{figure}

\begin{figure}
\includegraphics[width=84mm]{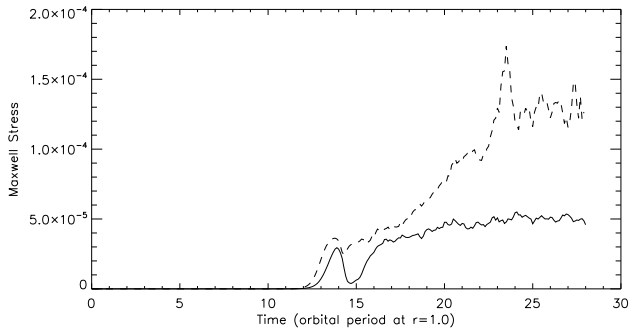}
\caption{Plot of the ideal MHD (solid line) and multifluid (dashed line) Maxwell stresses. 
}
\end{figure}

The Reynolds stresses show a similar evolution. However, the ratio of the ideal and multifluid Maxwell stress is larger than the ratio of the Reynolds stresses at saturation. This is expected as the stretching and twisting of the magnetic field and the resultant stress would be much more severe than the fluid motions that result from the fluids interaction with the magnetic field.

\begin{figure}
\includegraphics[width=84mm]{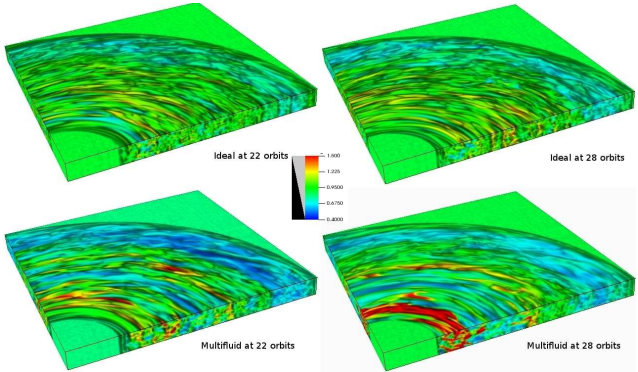}
\caption{A comparison of the neutral density at different times for both the ideal MHD and multifluid cases. The top two panels show the ideal neutral density at 22 orbital periods and 28 orbital periods respectively from let to right. The bottom two panels show the neutral density in the multifluid case at the same points in time. 
}
\end{figure}

Qualitatively, the turbulence that results from the MRI in the ideal MHD
regime appears to occur on smaller scales compared to the multifluid
regime (see Fig.9). Power spectra of the neutral density calculated from
the res4-ideal and res4-mf runs show that there is significantly more
power in all wavenumbers in the multifluid regime (see Fig.10). There is
a proportionately more power in the range $8<k<16$ in the full
multifluid case than in the ideal MHD regime. At large length scales
there is a significant difference in the power spectra showing that
larger scale structures are present in the multifluid regime.

\begin{figure}
\includegraphics[width=84mm]{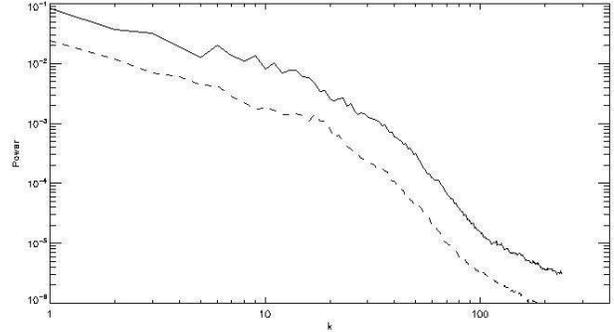}
\caption{Power spectra of the neutral density calculated from
the res4-ideal and res4-mf runs.
}
\end{figure}

\citet{Kunz:2013} describe large scale structures in the magnetic and
velocity fields, similar in nature to those found in the Hall dominated
multifluid model presented here. These structures are very long-lived
and have the net effect of reducing the amount of MRI driven turbulent
transport and slowing the flow of mass radially inward, whereas in the
res4-mf model the Hall effect leads to a net increase in turbulent
transport due to the MRI despite the appearance of these structures.
There are a number of important differences between the work presented
here and that of \cite{Kunz:2013}.  Likely to be of most significance is
that the results of \cite{Kunz:2013} are for {\em local} simulations and
therefore cannot reproduce the physics of large-scale (in comparison to
the radius at which the shearing box is located) structures in the disk.
Generally speaking, results from shearing box calculations which suggest
axisymmetric structures require global calculations to confirm their
validity as such structures may become unstable when allowed to evolve
in a global disk.  Secondly, the accretion disk in our work is fully
multifluid with self-consistent, spatially and temporally varying Hall,
ambipolar and parallel resistivities while \citet{Kunz:2013} utilise a
simplified description of the resistivities and include only Hall and
Ohmic diffusion.  Thirdly, in our work we perform isothermal
simulations, rather than the incompressible ones of \citet{Kunz:2013}.
Finally, the simulations presented in \citet{Kunz:2013} are run for a much
longer time period than those presented here, and the evolution of these
structures in the res4-mf model cannot be studied for a long time. It
is not known, therefore, whether their final state is different than that 
presented here. It is possible, nothwithstanding the likelihood of instabilities
reducing the impact of these structures on accretion, that they will continue 
to grow and eventually have a similar effect to that suggested in the
work of \citet{Kunz:2013}.

\subsubsection{Perturbed kinetic and total magnetic energy}

We find that the perturbed kinetic energy grows and saturates similarly to the stresses and $\alpha$-parameter for both the res4-ideal and res4-mf runs.

Surprisingly, the total magnetic energy is seen to grow exponentially
and it continues to grow strongly well after the time where all other 
parameters have saturated (see Fig.13). This is very surprising behaviour and 
does not have precedence in the literature. In a 2-dimensional non-ideal local
simulation \citep{Sano:2002a}, which looked at the non-linear evolution
of the MRI in the presence of the Hall effect when Ohmic resistivity is
not small. This behaviour seems to only occur when a net field is
applied to the computational box. This effect is attributed to the
channel solution. However, this phenomenon does not appear in the 3-dimensional simulations of \citet{Sano:2002b}. No evidence is seen in the res4-mf run to suggest that the channel solution is present.

The unbounded growth in the magnetic field is not a result of magnetic energy being introduced through the inner or outer boundaries. To ensure that this is the case, the total magnetic energy is simultaneously calculated for a number of annular regions throughout the disk and this is done throughout the life time of the simulation. The total magnetic energy is seen to grow exponentially in each annular region. The motions of the electron and ion fluids do not show accretion that is significantly different from the neutral species flow either. There is some build-up of magnetic energy immediately adjacent to the inner boundary but it is not seen to affect the dynamics outside this small region.

The power spectra for the components of the magnetic field are calculated, using a similar method to the azimuthal decomposition seen in the power spectra of \citep{Arlt:2001}, to determine if the unbounded field growth is more evident at small or large scales (see Fig.11). As seen in the ideal MHD case, the magnetic field is dominated by large scale components. This is expected with a magnetic instability such as the MRI which produces a large scale azimuthal field. As the disk evolves and the total magnetic energy increases exponentially and continues to grow, the power contained in each component grows on all wavenumbers. The azimuthal component still dominates over the others in the range $k < 10$, and this corresponds to the large scale azimuthal component. Interestingly, the growth of the power contained in this range stops increasing once the Maxwell stress saturates but the power contained at the smaller scales continues to grow.   

The difference between the components lessens at smaller scales and diverges at the smallest scales. At the smallest scales ($20 < k < 100$), i.e. the  region where the power seems to continue growing, especially in the azimuthal and radial components, where a very jagged profile is seen. However similar power spectra are calculated for the ideal MHD and ambipolar dominated runs and a similar feature are seen and so this is not specific to runs where the Hall effect is important. However, it is more pronounced in the multifluid MHD case.

\begin{figure}
\includegraphics[width=84mm]{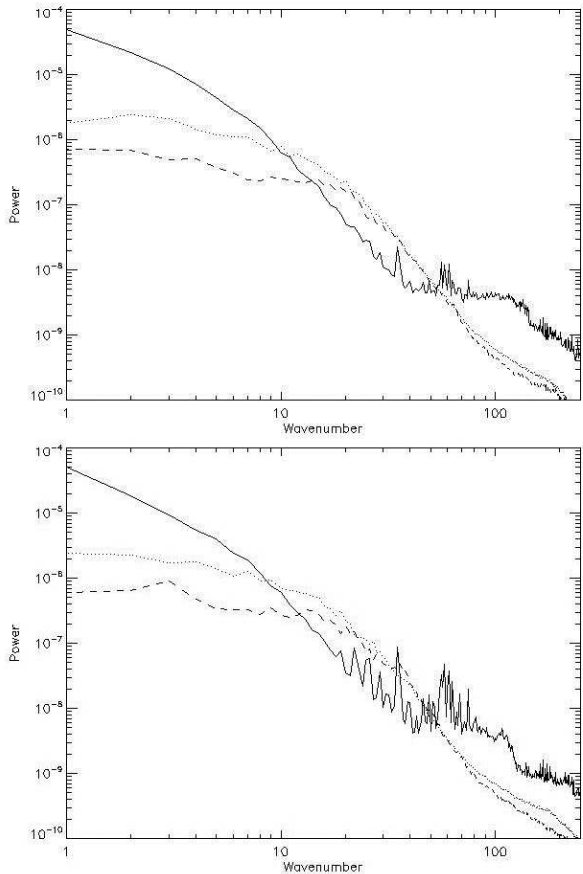}
\caption{Plot of the power spectra for the azimuthal (solid line), radial (dotted line) and vertical (dashed line) magnetic field components for the disk at 14 orbits (left pane) and 17 orbits (right pane) past perturbation injection.}
\end{figure}

It is likely that some small scale local process which is a consequence of the Hall effect is active. While the Hall effect is not a diffusive effect, it does reorient the magnetic field. This re-orientating of the magnetic field could give rise to a field configuration that compliments the local fluid flow so as to further destabilise the disk in addition to how it is already known to enhance the transport of angular momentum. This effect leads to the continuing growth of the magnetic field at smaller scales.        

It is clear from the results that multifluid effects have great importance in determining the properties of accretion disks in the parameter range studied here.

Another interesting effect seen in the res4-mf run (see Fig.9) is where gaps are seen to open in the disk. It is thought that gaps seen in accretion disks are thought to open due to planetary formation. Self-gravity is not modelled in this work and gap opening is due solely to non-uniform accretion.

\subsection{Effect of field orientation and net-flux}
Next we examine the effect of field orientation on the destabilisation mechanism of the Hall effect (res4-mf-minusbz). The numerical
set-up in this case is exactly the same as the res4-mf run except that the initial magnetic field is in the direction $-B_z$ such that
the sign of $\boldmath{\Omega} \cdot \mathbfit{B}$ is negative. In this regime, the Hall effect is actually a stabilising mechanism in the disk \citep{Wardle:2012}. 

\begin{figure}
\includegraphics[width=84mm]{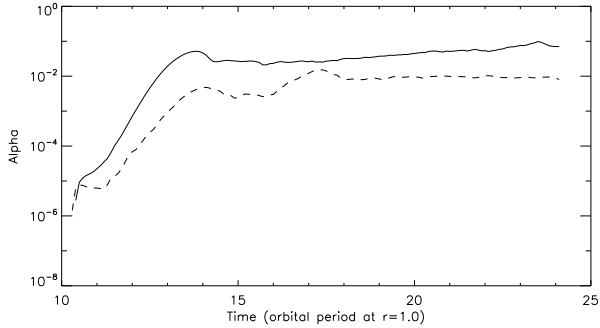}
\caption{Log-linear plot of the $\alpha$-parameter for the res4-mf (solid line) and 480-mf-minusbz (dashed line) runs.
}
\end{figure}

We find that in this case the MRI is heavily suppressed although some
growth is seen and weak turbulence is present in the disk. Saturation in the diagnostic parameters occurs early and at relatively smaller amplitudes compared to the res4-mf case (see Table 1). For example the Maxwell stress and the $\alpha$-parameter saturate at roughly 10\% of the value seen in the res4-mf run (see Fig.12). An order of magnitude drop in these parameters is significant and the consequence is that angular momentum transport occurs at a much reduced rate. 

It is seen that the total magnetic energy does not show the same properties as seen in the res4-mf case where it grew exponentially without bound (see Fig.13).

\begin{figure}
\includegraphics[width=84mm]{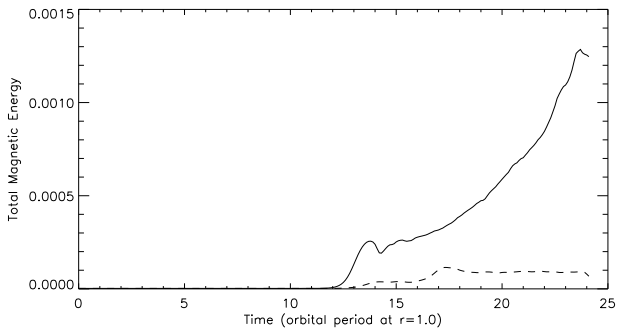}
\caption{Plot of the total magnetic energy for the res4-mf (solid line) and the 480-mf-minusbz (dashed line) runs.
}
\end{figure}

Next, a run with no net magnetic flux is performed. In the paper of \citet{Hawley:1992}, 2-dimensional local simulations are carried out in the ideal MHD framework which showed that the growth rate and saturation amplitude of the MRI are directly affected by the initial flux, increasing as the net flux is increased. In the case of a zero net flux, there is an initial growth period followed by saturation and subsequent decay of the turbulence due to the anti-dynamo theorem of \citet{Moffatt:1978}. However, our simulations do not satisfy the conditions set out in this paper due to our system not being an isolated one.

\begin{figure}
\includegraphics[width=84mm]{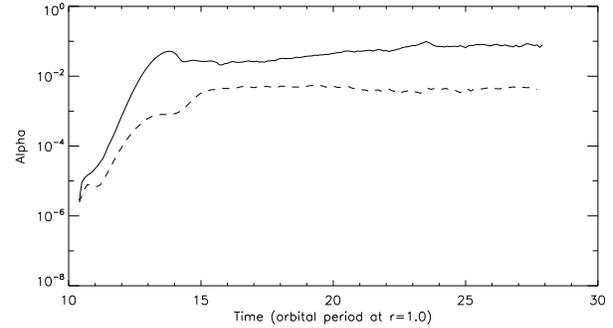}
\caption{Plot of the $\alpha$-parameter for the res4-mf (solid line) and the 480-mf-zeronet (dashed line) runs.
}
\end{figure}

The magnetic field is set up using the following configuration,

\begin{equation}
B_z(r)= B_0 sin \left ( 2 n \pi \frac{(r - r_{i})}{(r_{o} - r_{i})} \right )
\end{equation}

\noindent and the radial boundaries are adjusted to take this into account.

In this case the $\alpha$-parameter and Maxwell stress saturate at roughly 6\% of those calculated in the res4-mf run. Again the total magnetic energy shows very weak growth and saturates at an early stage in the development of the disk once any large local regions of positive net flux have mixed out, the destabilising hall effect becomes weak and no longer plays a role. A very slight decay is seen in all diagnostic parameters for the remaining life time of the disk.

A physical situation where the disk has no net magnetic flux is probably an unrealistic one and it is reasonable to assume that large regions of the disk have at least a weak net magnetic flux.

\begin{figure}
\includegraphics[width=84mm]{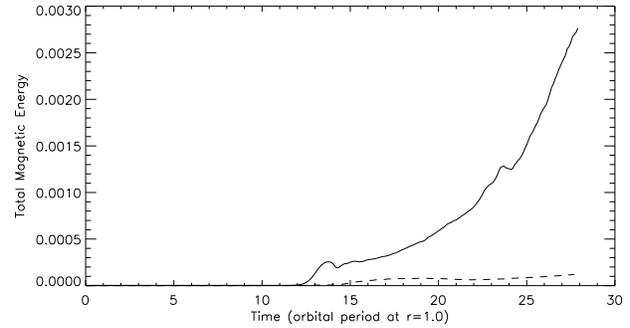}
\caption{Plot of the total magnetic energy for the res4-mf (solid line) and the 480-mf-zeronet (dashed line) runs.
}
\end{figure}

The results presented in this section confirm and validate those of
\citet{Sano:2002a,Sano:2002b} for a global accretion disk in the
multifluid regime where Hall is initially the dominant non-ideal effect.
 
\section{Conclusions}

In this, the first of two papers, we have presented the results of a series of simulations designed to examine whether non-ideal effects manifested by the multifluid regime affect the long term dynamics of accretion in disks around protostars. A weakly ionised disk model is implemented in a region of the disk where the Hall effect is believed to dominate over ambipolar diffusion.

The results of this study show, that angular momentum transport, parametrised by the $\alpha$-parameter, is significantly enhanced by inclusion of all the non-ideal effects in this parameter space. The most surprising result is that in the multifluid regime such an accretion disk, active to the MRI, produces an unbounded exponential growth in the total magnetic energy. It is not clear if saturation would occur at a later time, though it may be assumed that once the magnetic energy reaches high enough levels the MRI would cease to be active as the disk would effectively lock up.

The case where $\Omega \cdot \mathbfit{B}$ is negative leads to a suppression of the MRI and much slower rates of angular momentum transport. Likewise for the zero net field case, but for different reasons. These results confirm similar results found in the literature.

Our results strongly suggest that the Hall effect is responsible for
enhancement of the MRI where a net field (with appropriate orientation)
is present. In a forthcoming paper the individual effects of ambipolar 
diffusion and the Hall effect on the dynamics of a protoplanetary disk will be 
studied. 

\section*{Acknowledgements}

This research was part-funded by Science Federation Ireland. The authors wish to acknowledge SCI-SYM for use of the HEA computational cluster in D.C.U. We acknowledge the PRACE for awarding us access to resource JUGENE based in Germany at the J\"{u}lich Supercomputing Centre.




\label{lastpage}

\end{document}